\def\folio{\ifnum\pageno<2\nopagenumbers\else\number\pageno\fi}
\newtoks\headline \headline={\hss\twelverm\folio\hss} 
\newtoks\footline \footline={{\hfil}} 
\font\mathbf=cmmib10 scaled 1000             

\def\ref{\par\noindent\hangindent=2pc \hangafter=1 }

\def\cappage #1 #2 #3 {\vfill\eject\pageno=#1 
\vglue 10 true in minus 10 true in \noindent{\bf Figure #2.} #3}
\def\ee #1 {\times 10^{#1}}
\def\ut #1 #2 { \, \hbox{#1}^{#2}}
\def\u #1 { \, \hbox{#1}}

\def\msol{\, \hbox{$\hbox{M}_\odot$}}
\def\kms {\, \hbox{km}\,\hbox{s}^{-1}}

\let\grad=\nabla
\def\cross{{\bf \times}}
\def\curl #1 {\grad \cross #1}
\def\div #1 {\grad \cdot #1}

\def\msol   {\hbox{$M_\odot$}}                  
\def\kms    {\hbox{km{\hskip0.1em}s$^{-1}$}}    

\documentstyle[11pt,paspconf,epsf]{article}

\begin{document}

\title{The Interaction of two Prominent 
Galactic Center Sources: Sgr A East and the Molecular Ring}
\author{F. Yusef-Zadeh}
\affil{Northwestern University, Evanston, IL. 60208}

\author{S. Stolovy}
\affil{University of Arizona, Tucson, AZ 85721}

\author{M. Burton}
\affil{University of New South Wales, Sydney, NSW, Australia}

\author{M. Wardle}
\affil{University of Sydney, Sydney, NSW, Australia}

\author{F. Melia}
\affil{University of Arizona, Tucson, AZ 85721}

\author{J. Lazio and N. Kassim}
\affil{Naval Research Lab., Washington, DC 20375-5351}

\author{D.A.  Roberts}
\affil{University of Illinois, Urbana, IL 61801}

\begin{abstract}

We present a synthesis of 
a number of 
recent observations in the near--IR H$_2$ and [Fe\,II] lines, OH (1720 MHz) 
maser line  and 
various radio continuum measurements using
the NICMOS of the HST, UNSWIRF on the 
AAT and the VLA. 
These observations suggest that the outer edge of the CND 
is collisionally excited 
whereas the inner edge is
likely to be heated predominantly 
by the IRS 16 cluster. The velocity and spatial
correlation of H$_2$ and OH (1720 MHz) as well as the spatial distribution
of radio continuum emission at 90cm  suggest that Sgr A East is
 responsible for shocking the gas and 
interacting with  the circumnuclear ring at the Galactic center. 
\end{abstract}

\keywords{galaxies: ISM -- Galaxy: center -- ISM: Sagittarius A East, West}

\section{Introduction}

The picture that has emerged from a suite of multi-wavelength observations of 
the Galactic center is that it contains a rotating, clumpy 
molecular ring (also known as the Circumnuclear
Disk- CND) which is heated by a centrally concentrated source 
(the IRS 16 cluster) of UV radiation.  Within the ring's central cavity, 
three ``Arms'' of ionized gas (Sgr A 
West) are generally in orbital motion around the center, which is believed to
contain a $\sim$ 2.5$\times$10$^{6}$ \msol  black hole (Ekart and Genzel 1996; Ghez et al. 
1998).
The coincidence in the geometry and kinematics of
the southwestern edge of the molecular ring and ionized gas 
suggests that  the  ionized gas is  
dynamically coupled to the inner edge of 
 the circumnuclear ring. The   western edge of the ring 
(sometimes  called the Western Arc) is 
out of the plane of the sky and obscuring the background 
stars and absorbing 
Br $\gamma$ emission from the western arm causing  the ratio of
H92$\alpha$ to Br $\gamma$ to increase (see recent reviews by Genzel, Hollenbach and 
Townes 1994; Morris and Serabyn 1997). 
Although the CND is often described as having 
a  circular geometry and kinematics, 
(e.g. Morris and Serabyn 1997 and the references cited therein), the ring structure shows
deviations from circular geometry. A 
prominent gap in the distribution of
molecular gas  in the CND shows  non-circular moving gas (e.g. Jackson et al. 1993). 
The CND gap is located 
where  an  ionized ``streamer'' appears to run along    
the extension of the N.  Arm beyond the CND. The thermal ionized gas in
the streamers show that the Northern  Arm of Sgr A West is  not coupled
dynamically 
and spatially to the
molecular gas in the inner  edge of the CND. 

On a larger scale the non-thermal structure known as Sgr A East is thought to be a
shell-type explosive event, possibly a  supernova remnant (SNR),  surrounding
 the thermal source Sgr A West. 
 The recent discovery of
OH(1720MHz) masers  at the interface of the 50 \kms\ molecular cloud and Sgr A East
provides strong evidence that these two are physically interacting
with each other (e.g. Mezger et al. 1996; Zylka et al. 1992; Serabyn et
al. 1992; 
 Yusef-Zadeh et al. 1996).  
The masers are detected to the southeast of 
Sgr A East (sources A, D-G)
as well as   
to the  northwest of the  CND.  The maser spots
surrounding the Sgr A East shell have radial velocities
 near 50 \kms\ close to the systemic 
velocity of the SNR.  The
expansion of a supernova into the Sgr A East molecular cloud (i.e.
the 50 \kms\ cloud) is  considered to be responsible
for the production of  the shocked OH(1720 MHz) maser emission.

The nature of the association  between the two systems,
Sgr A East and Sgr A West and their corresponding molecular clouds, 
  is  not well established. 
Radio continuum studies have   pointed out
the curious alignment of the western 
edge of Sgr A East and the Western Arc of Sgr A West
(Ekers et al. 1983; Yusef-Zadeh and Morris 1987; Pedlar et al. 1989). 
Sgr A East is   
 behind Sgr A West, a conclusion  based on low-frequency observations 
showing a hole in the 
distribution of ionized gas toward the Arms of Sgr A West
 (Yusef-Zadeh and Morris 1987; Pedlar et al. 1989).
 These observations imply that  
a segment of the  Sgr A East shell must lie behind 
the Galactic center by a distance that is estimated to
range between a few to hundreds of parsecs. 

The relative location of Sgr A East with respect to the Galactic center 
is of interest for a number of reasons. First, the separation between the two
 systems  is relevant to the role that the relativistic particles of 
Sgr A East may play in
upscattering  the CND's IR and UV photons to high energies.
  The strength  of this interaction depends 
on the distance between Sgr A East and the CND. This process  
was recently incorporated into the modeling the high-energy emission from 
the Galactic center based on EGRET measurements (Melia et al. 1998a,b; Fatuzzo 
et al. 1998). 
In this model, Sgr A East is considered to be within 5pc of the 
Galactic center.  Second,  the disturbance of the   molecular
ring  by Sgr A East also  depends upon the separation between the 
two objects. 
Recent studies  suggest that the N. Arm of Sgr A West 
may  delineate the inner edge of  
 an infalling ``tongue''  of neutral gas  toward the Galactic center.
The obvious question that needs to be addressed is whether  
Sgr A East is  responsible for the observed deviation from circular motion 
at the location of the gap in  the CND and for the infalling cloud. 
Lastly, the closeness of  Sgr A East to the Galactic center and its unusual energetics 
brings up the possibility that the sources  responsible for the explosion are 
 the family of young, massive stars in the IRS 16 cluster.
 This hypothesis may provide an insight into the 
nature of massive star formation within the same parent cloud in the inner
5pc of the Galactic center. 

A signature of an interaction between Sgr A East and the CND 
should be shocked gas as the blast wave of Sgr A East compresses the gas 
in the  CND. 
The unambiguous detection of shocked (as opposed to UV-heated) gas associated with 
the Galactic center has been difficult.  The
intensity ratios of the v=2--1 and 1--0 S(1) lines of H$_2$ (Gatley et al.
1984;  Burton \& Allen 1992; Pak et al.  1996) are often
consistent with collisional excitation rather than fluorescence, 
but the high density of the molecular gas at the Galactic center
and the intense UV radiation field in the region allow the line 
ratios from UV-irradiated gas to resemble those of shock-heated gas
(Sternberg \& Dalgarno 1989). However, the presence of the OH 1720 MHz line, 
and the absence of the 1665/1667 MHz lines provides a
clear diagnostic of shocked molecular gas, as the FIR 
radiation field from warm dust in UV-heated clouds would pump the latter
transitions. Here we report preliminary  results of our 
new  H$_2$ and [FeII] and 1.2 and 90cm radio continuum observations
and examine the interaction picture of Sgr A East  with the CND. 

\section{Observations}

We observed the
Galactic center in the  
 1--0 S(1) transition 
of H$_2$ line using both the NICMOS on the HST
as well  UNSWIRF on the AAT. 
The UNSWIRF observations were carried 
out using four overlapping frames  with a size of 
90$''$, a pixel size of 0.77$''$
and  channel maps 
 separated  by $\approx$39  \kms\ (Figs. 1,2 and 3). 
Using camera 3 of NICMOS, a total of  12 overlapping frames each 
with a 52$''$ size and a pixel size of 0.203$''$
was observed in both the line and the continuum.
Each frame was dithered by 
16$''$ in each  principal directions of 
the detector,  and the final image was constructed by 
mosaicing   all 12 
frames, a portion of which  is shown in Figure 2.  
The 90cm  continuum observations were carried out in 
the BnA (Fig. 4) and A-array configurations (Figs. 5 and 6)
of 
the Very Large Array of the National 
Radio Astronomy Observatory\footnote{The National Radio Astronomy 
Observatory is a facility of the National Science Foundation, operated 
under a cooperative agreement by Associated Universities, Inc.}.   
Camera  3 of NICMOS  was also used to observe the [FeII] line emission
from the central 110$''\times160''$, a subsection of which is shown in Figure  8b.
The NICMOS images were produced by subtracting images in the 1\% filter
centered on the 1.64$\mu$m  [FeII] line. The continuum was adjusted 
to minimize the stellar residuals, but both positive and 
negative residuals remain. Some of the residuals are due to differences 
in intrinsic colors and locally patchy extinction. 
These observations have provided 
wealth of structural details with subarcsecond resolution capability of 
the NICMOS, the velocity  information from UNSWIRF, and  the 
ability to discriminate line from continuum emission that UNSWIRF gives.

The high resolution 1.2cm image were based on combined A and B array configurations 
whereas the  low-resolution image was based on C configuration. 
  The astrometry of the radio and IR 
images was done by using the position  of IRS 7 with respect to Sgr A$^*$ 
(Menten et al. 1995). A detailed account of
these
observations will be given elsewhere. 
 

\section{Results}

\subsection{H$_2$ Filament and H$_2$ Gas inside the CND}

Figure 1 shows contours of 1--0 S(1) H$_2$ line emission 
measured by UNSWIRF superposed on the greyscale 6cm 
continuum emission from Sgr A West and East (Yusef-Zadeh and Morris 1987). 
A segment of 
the Sgr A East nonthermal shell is seen to the NE near 
$\alpha = 17^h 42^m 32^s,  \delta = -28^0 58'$ 
whereas
the spiral-shaped structure of Sgr A West is evident  near the
 center of the Figure. 
The peaks 
located in the 
NE (N. lobe) and SW (S. lobe) parts of the
 ring are
consistent with limb-brightening of the inner edges along the principal
axis. 
An extended H$_2$ feature with a plume-like appearance 
is noted within the ring. This new feature 
appears to terminate at the position of IRS 7 (a more detailed account of
this feature will be given elsewhere). 
The distribution of H$_2$ line emission appears like a ring 
surrounding Sgr A West except 
in the region near $\alpha = 17^h 42^m 30^s,  \delta = -28^0 58' 45''$, 
where
the HCN distribution  indicates 
non-circular motion.
 at the location of the inner 
gap (Jackson et al. 1993).  
A long
filamentary structure with an extent of $\approx1'$ is
discovered to the NW of the CND.  A    gap is noticed along
the extent of the  filament near 
$\alpha = 17^h 42^m 29^s,  \delta = -28^0 58' 30''$ 
which  shows a similarity in its spatial distribution 
to the iner  gap of the CND.
 The crosses in all the figures show the positions of OH (1720MHz) masers 
and the star symbol is coincident with the position of Sgr A$^*$. 
The C and B masers with    velocities  of 43 and 134  \kms\ 
are seen near the outer and inner gaps of H$_2$ distribution.

\begin{figure}
\caption{Contours of H$_2$ 1--0 S(1) emission from the CND
with a pixel size of 0.77$''$ is superposed 
on a 6cm radio continuum image of thermal emission from Sgr A West 
with a spatial resolution of 0.67$''\times0.4''$.  
 A segment of the 
nonthermal Sgr A East shell is seen to the NE. The star symbol represents Sgr~A$^*$
and the crosses show the position of OH (1720 MHz) masers.} 
\end{figure}

Figure 2 shows
the same contours as in Figure 1, but  they are now superposed 
on the greyscale H$_2$ image based on the NICMOS observations. 
Because of the complex nature of Galactic center sources, 
the stellar continuum in NICMOS images 
has not been removed completely. 
There is an overall agreement between the 
NICMOS and UNSWIRF data  in spite of the contamination
from   
stellar continuum in NICMOS images.  
The new H$_2$ filament appears strikingly sharp with a width of
about 0.5$''$ in the  NICMOS image. 

\begin{figure}
\caption{The same as Figure 1 except that the greyscale image is replaced by
the  HST/NICMOS H$_2$ image. The greyscale image is convolved by a Gaussian 
having a size twice the 
pixel size of  0.203$''$. } 
\end{figure}

 Figure 3 shows the velocity distribution of H$_2$ 
emission based on one 90$''$ field 
 centered near the N.  lobe as observed with UNSWIRF.  
 The crosses 
indicate the position of the masers. 
 The N. lobe of the CND 
appears  prominantly at higher velocities near the 120 and 
 159 \kms\   channel maps. 
The new H$_2$ filament appears as a coherent velocity 
feature  and is best seen  in the  41.2 and 80.4 \kms
channel maps.
The C maser with a velocity of 43 \kms\ is probably related to the new H$_2$ 
filament, whereas the B masers,
with velocities of 
about 134 \kms\,  are associated with the N. lobe. 
The overall velocity 
structure of the new filament is also similar  to that of the
 inner edge of the CND near the Western Arc,  suggesting that the new 
filament  is part of the CND but 
outlining the outer  edge of the CND. A ridge of weak H$_2$ emission 
is also  seen  in the 41 and 80 \kms\ channel maps. This feature 
appears to connect the two gaps, as
 seen in   
the new filament 
and in the inner edge of the CND,  suggesting that the H$_2$ features are indeed 
associated with each other.

\begin{figure}
\caption{The H$_2$ velocity distribution of a field 
 centered on the 
N.  lobe of the CND is shown in reverse greyscale. The LSR channel velocities 
are  shown on top right corners and the crosses represent OH maser sources, C to the 
right and B to the left. The high
and low values are shown in black and white,
respectively. The white features are 
artefacts from imperfect continuum subtraction. } 
\end{figure}

\subsection{Radio Continuum  Features at 90 and 1.2cm} 

Figure 4  shows the greyscale distribution of 90cm emission 
superposed on contours of H$_2$ emission. 
The hole at the center is 
due to  free-free absorption of the arms of Sgr A West  against Sgr A East. 
The 90cm  image 
also shows the ionized streamers in absorption at the position of the 
inner and outer gaps.  
The elongated  structure of Sgr A East,  which  is 
saturated in this image,  delineates the bright shell of Sgr A East. 
The western  edge of the shell is weaker in its surface 
brightness due to optical depth effect but what is clear is a deviation 
from a shell-like geometry to the west of the CND. 
We also note that the CND is clearly surrounded by 
nonthermal emission (Pedlar et al. 1989) not only 
to its  western edge   but also  to its eastern edge. The prominent eastern shell of Sgr A East 
lies near $\alpha = 17^h 42^m 38^s$ which is about 1.5$'$ east of the 
eastern edge of the CND. Figure 5 presents a more close up view of the
90cm  emission with a higher resolution
where contours of HCN emission (Wright et al. 1987)
appear to be surrounded by Sgr A East.  
In addition,  the 90cm emission   
shows an elongated weak nonthermal feature within  the hole. 
The striking feature in this image is the continuum emission 
crossing  the bar of ionized gas
associated with Sgr A West.
This shows clearly a  lack of  free-free absorption of a segment of 
the bar against 
Sgr A East.

Figure 6 shows contours of 90cm  emission superposed on a greyscale 1.2cm 
continuum image of Sgr A West showing the extent of the elongated low-frequency 
feature with respect to Sgr A West and Sgr A$^*$. 
The presence of low-frequency 90cm emission from the direction of
the bar and Sgr A$^*$ is  surprising because the
 bar of ionized gas is known to be the
 brightest feature with the highest ionized density in   Sgr A West.
If Sgr A East lies behind Sgr A West, the bar should have 
the  largest  optical depth at 90cm. 
This suggests that the Sgr A East shell is 
not totally behind Sgr A West as the elongated feature
which is probably a nonthermal feature associated with Sgr A East
 crosses 
the hole created by free-free absorption. In fact, the comparison
of the molecular and ionized material with the 90cm image, 
as presented in Figures 5 and 6, gives the impression of 
an   anti-correlation between the molecular material and the 
90cm  emission. The lack of nonthermal emission from Sgr A East
versus the free-free absorption of thermal Sgr A West against 
Sgr A East will be considered in more detail elsewhere. 

Radio continuum emission  at $\lambda$1.2cm traces 
  the distribution of thermal gas  in 
Sgr A West and is unlikely to be  contaminated by the nonthermal Sgr A East 
given its steep radio  spectrum in radio wavelengths (Pedlar et al. 1989). 
Figure 7 shows contours of 
HCN emission superposed against the 1.2cm greyscale of Sgr A West. 
This figure shows the prominent three arms of Sgr A West and a  
faint Arm of ionized gas which appears to  delineate the outer western edge 
of the CND. 
This new and extended ionized feature runs between 
$\alpha = 17^h 42^m 27^s,  \delta = -28^0 59' 50''$ 
and $\alpha = 17^h 42^m 28^s,  \delta = -28^0 59' 15''$
and has  a typical brightness  between 0.5 to 1.5
mJy/beam 
with a beam size of 1.58''$\times0.76''$. There is also a peak  
of  90cm emission  coincident with the new high-frequency 1.2cm
feature near 
$\alpha = 17^h 42^m 26.5^s,  \delta = -28^0 59' 30''$ 
with a flux density of 80 mJy in a beam of $9.5''\times5.1''$. 
It appears that  the western edge of the CND
is sandwiched by two ionized layers from east and west along   the inner and outer 
edges of the CND, respectively. 
The eastern  or the inner layer of the CND is traced by
the S. arm (i.e. the Western Arc) 
and is ionized by the strong UV radiation 
field originated by the IRS 16 cluster. 
The western or the outer layer of ionized gas is  identified as a new and weakly emitting 
arm  of ionized gas tracing the outer edge of the CND. The latter     can  either 
be due to photoionization  by the radiation field from nearby stars 
lying outside the CND or by shocked gas as a result of an interaction 
with Sgr A East as  a  J-type shock is driven into the CND. 

\begin{figure}
\caption{ Contours of H$_2$  distribution superimposed on the 90cm emission
from Sgr A East 
with a spatial resolution of 16.5$''\times10''$.  
Sgr A West is absorbed against Sgr A East at this low frequency.}
\end{figure}

\begin{figure}
\caption{Contours of HCN emission (G\"usten et al. 1987; Wright et al. 1987) 
with a spatial resolution of 5-7$''$  
superposed on the 
central region of the 90cm distribution 
with a spatial resolution of 9.5$''\times5.1''$.}
\end{figure}

\begin{figure}
\caption{Contours of 90cm emission as seen in 
figure 5 are superposed on 
the greyscale distribution of 1.2cm emission
with a spatial resolution of 1.6$''\times0.8''$. The star marks the 
position of Sgr A$^*$.}
\end{figure}

\begin{figure}
\caption{A greyscale distribution of ionized gas at 1.2cm
as shown in figure 6 
superimposed on contours of  HCN distribution}
\end{figure}

\subsection{Dark Patches} 

 We  note prominent 
dark patches 
distributed throughout the H$_2$  images. These dark features
which are  best seen in  NICMOS images in Figure 2
  are extended and appear to 
be concentrated prominantly along 
the inner and outer gaps of the CND and 
along  the 
remarkably extended structure NW of the H$_2$ filament. 
Although the lack of proper removal of stellar continuum  
contributes  to the absorption features in  NICMOS images, all the 
features noted here are extended,  have been detected in continuum 
images and show
atomic or 
molecular couterparts. Estimates of the extiction toward these
patches will be given elsewhere. 

An  example of the dark feature in Figure 2 is 
  located near the N.  lobe but to the south of the 
inner gap of the CND   near
$\alpha = 17^h 42^m 30^s,  \delta = -28^0 58' 50''$. 
This feature  coincides with  the well-known  OI and
dust feature  
within the CND (Davidson et al 1992; Jackson et al. 1993; Telesco et al. 
1996). The OI feature is thought to lie 
in the CND with H$_2$ density  of 3$\times10^5$ cm$^{-3}$ 
 and a mass of 300 \msol
 suggesting 
that the dark feature in this region is located at the Galactic center. 
The  strongest   support for this suggestion 
comes from the distribution
of radio continuum
emission at 1.2cm 
and the distribution of [FeII] feature based on the VLA and HST/NICMOS
measurements. 
Figure 8a shows the N.  Arm of Sgr A West and a faint  semi-circular shell
structure surrounding the northern  tip of the N. Arm. This weak 
shell-like ionized feature with a typical brightness  of 
0.2  mJy within a beam of 0.3$''\times0.2''$ coincides with the outer rim of 
the OI/dust  feature and the dark H$_2$ patch. 
 This shell of ionized feature suggests that the 
the dark patch  is photoionized externally.


 Figure 8b  shows the 
distribution of [FeII] in the same region as shown in Figure 8a.
A   dearth  of [FeII] emission  corresponding to the 
northern half of the N. Arm is seen at the position of H$_2$ 
dark patch. Both the [FeII] and H$_2$  features are 
surrounded by a rim of ionized gas in the  
1.2cm image. 
We note a  lack of correlation along the Northern arm 
in 1.2cm and [FeII] images,
as shown in figures 8a and b. 
The  northern half of the N.  Arm also shows 
 a sudden drop by a factor of 2  in 
its brightness  to the north of 
$\alpha = 17^h 42^m 29.7^s,  \delta = -28^0 59'$ (Yusef-Zadeh and Wardle 1993). 
Considering that free-free emission at 1.2cm is optically thin, the drop in the 
emission measure as the Northern arm crosses the shell 
is a strong support for the association of the OI/dust feature with
the N. arm. 
%

\begin{figure}
\caption{(left) A greyscale distribution of ionized gas 
at 1.2cm with a resolution of 0.3$''\times0.2''$
 showing the Northern arm and 
a shell of weak ionized gas around its northern tip.
(right) A greyscale distribution of 
[FeII] ionized gas showing a lack of emission centered on
the northern half of the N. arm. }
\end{figure}


The coincidence of
OI/dust  emitting feature with  the dark features and 
an    ionized rim  
surrounding them  
suggests that they   are all located within the CND. 
Further support comes from 
a drop in the emission measure of optically thin free-free emission
from the N. Arm at 1.2cm.    A detailed examination of the 
ionized gas in Sgr A West shows also a number of dramatic dark features 
which can not be explained by a  lack of short-spacing  data. These
dark continuum features are likely to delineate the region of 
cold dust and gas where  the emission measure drops significantly.
Recent proper motion results show the direction of the ionized flow
supporting the infall picture of the N. Arm toward the Galactic center
(Yusef-Zadeh, Roberts and Biretta 1998; Zhao and Goss 1998). The large-scale
distribution of dark  features in the H$_2$ images suggests that the 50 \kms\ 
could be responsible for the infall of neutral material toward the Galactic center
as evidenced by the disturbed molecular material  in the  CND at the location of the
inner and outer gaps. 
The location of the dark patches along the N. Arm, near 
Sgr A$^*$, suggests that the atomic and molecular gas 
 survives well within the cavity of ionized gas.
 
We also note  the darkest feature lying  close to the 
IRS 16 cluster with very bright continuum emission. Because of 
the strong stellar emission from the IRS 16 cluster, this dark feature may be 
due to 
artefacts from imperfect continuum subtraction.  
However, it is interesting 
to  note that the dark feature  coincides 
with the highly blue shifted 
H$_2$CO, OH, HI  and HCO$^+$ absorption 
features,  with velocities of about --190 \kms\.
(Marr et al. 1993; Pauls et al. 1993;  Yusef-Zadeh et al.1993;
Yusef-Zadeh 1994; Zhao et al. 1995). The  kinematic and spatial distribution of the
absorbing gas suggest that 
the gas  is  located 
at the Galactic center
(see Liszt and Burton 1995 for an alternative interpretation). 



\section{Discussion}

The new H$_2$ filament  
running   parallel to the nonthermal shell of Sgr A East 
supports the picture that the impact of the nonthermal shell of
Sgr A East with a molecular cloud is responsible 
for shocked H$_2$ emission from the filament. 
Additional support for the shock hypothesis comes from  the sheet-like 
morphology and the location of maser source C 
 along the filament. The radial velocity of the 
C maser is about 43 \kms\ which is within the 
H$_2$ velocity range of velocities of  40 and 80 \kms\,
as shown in Figure 3. These so-called supernova masers are 
thought to be excellent tracers of
shocks and are seen primarily in regions where 
SNR's are interacting with molecular clouds (Wardle et al. 1998; 
Green et al. 1998; Frail et al. 1996). The lack of any thermal continuum 
or Br $\gamma$ emission at the location of the new filament  is consistent 
with the 
C-type shocks being  responsible for exciting the OH gas (Wardle et al. 1998). 

The compelling evidence that the new filament traces the shocked molecular
 gas by the expansion of the Sgr A East shell into the
cloud raises the question of the location of the filamentary shocked gas with
 respect to the CND. It is possible that either the
shocked gas is associated with the 50 \kms\ cloud and it happens to be aligned 
fortuitously along the outer edge of the CND or that the
outer edge of the CND is shocked externally as a result of its interaction 
with Sgr A East.  What follows is a series of arguments
in support of 
the outer boundary of the CND being collisionally
 excited by the
 expansion of the nonthermal Sgr A East
into the CND. 

First,  
the anti-correlation of the streamers of
ionized gas  and molecular  gas  at the positions of the inner and outer 
H$_2$ gaps 
show  compelling morphological evidence for the association of 
 the streamers of Sgr A West and the new H$_2$ filament which  
traces  the shocked gas associated with the 50 \kms\ molecular cloud. 
Second, 
the inner and outer boundary where the gaps are observed show similar 
velocity distributions suggesting that 
 they both must be part of the molecular ring. 
A  ridge of H$_2$
 emission linking the new filament 
to the inner edge of the CND furthers the support for the physical 
association between them. 

Third, the evidence that Sgr A East and the CND are physically 
interacting with each other comes from detailed 90cm 
observations 
of the Galactic center. 
In these  images, the western and eastern edges of 
the CND appear to be
engulfed  by a shell of radio continuum emission at 90cm. 
 The emission 
at these low-frequencies  are due to nonthermal processes and 
is expected to be  associated with Sgr A East.
In addition, the evidence for low-frequency emission from the region of the 
bar and absorption of the arms of Sgr A West add further 
evidence that  Sgr A East surrounds  the CND. 
The high frequency 
radio emission delineating the outer edge of the Western Arc of the CND 
could also be used as an argument for the interaction hypothesis if the
radiation field outside the CND is not strong enough to ionize the outer 
edge of the CND. A lack of [FeII] emission from the northern half of 
the N. arm and its correlation with
OI/dust  emission and H$_2$ dark patches as well as a number of kinematically 
disturbed features in the CND all could be used as implications of 
a scenario in which a cold dust cloud plunges into the CND by an energetic 
phenomena.

Lastly, the recombination line emission toward Sgr A East and 
Sgr A$^*$  show stimulated thermal gas at velocities near 50 \kms\ (Anantharamaiah et al. 
1998; Roberts and Goss 1993). It is possible
that the 
thermal ionized gas is associated with the 50 \kms\ molecular cloud. If so, 
a part of the 50 \kms\ molecular cloud with which Sgr A East is interacting
must be in front the of Galactic center.

\section{Summary}

The conclusion that 
Sgr A East and the CND are interacting is inescapable based on 
various lines of evidence and arguments given above. 
A further test of  
this model could be done by  
measuring  the intensity ratio of v=2--1 and 1--0 S(1) line emission from the 
region  already observed  presented   in v=1--0 S(1) line. 
The implication of such an interaction is that the inverse Compton 
scattering of IR and UV photons should not be ignored in modeling 
the high-energy 
emission from the Galactic center. Furthermore, the disturbance of the 
gas in the inner edge of the CND is caused by the impact of Sgr A East.  
These suggest that
 Sgr A East is responsible  for pushing  
 gas cloud into the 
cavity. In this scenario,  the N. Arm traces the
trajectory of the infalling cloud as it is  ionized by the strong 
UV radiation field of the IRS 16 cluster. The three-dimensional motion 
of ionized gas based on proper motion and radial velocity 
measurements (Yusef-Zadeh, Roberts and Biretta 1998; Zhao and Goss 1998)
is consistent with this picture. 
The dark patches seen within the inner 30$''$ of Sgr A$^*$, 
in the CND coincident with OI feature, in the inner and outer gaps and the 
extended dark feature to the NW of the filament in NICMOS image of H$_2$ distribution 
gives the impression that the 50 \kms\ molecular cloud may in fact be 
responsible for feeding the Galactic center (see Coil and Ho 1998).  
Lastly, this interaction
picture of Sgr A East and the CND may 
shed some new light on the possibility 
of the common origin of the IRS 16 cluster and the 
progenitor of stars 
responsible for the formation of Sgr A East.

\acknowledgments{ We thank Mel Wright for providing the HCN image and 
M. Ashley, S. Ryder and J. Storey for assistance with using UNSWIRF. 
Basic research in
radio astronomy at the Naval Research Laboratory is supported by the
Office of Naval Research.
TJWL is
supported by an NRL-National Research Council Associateship.}


\begin{references}

\ref Anantharamaiah, K.R. et al. 1998, this conference. 

\ref Burton, M. \& Allen, D.A. 1992,
 in {Proc. Astr. Soc. Australia},  10, 55

\ref Coil, A.L. and Ho, P.T.P. 1998, ApJ, in press.

\ref Davidson, J.,   et al. 1992, ApJ, 387, 189




\ref Ekart, A. \& Genzel, R. 1998, this conference.

\ref Ekers, R.D., van Gorkom, J.H., Schwarz, U.J. \& Goss, W.M.
1983, A.A. 122, 143

\ref Fatuzzo, M. et al.   1998, this conference.

\ref Frail, D. A., Goss, W. M., Reynoso, E. M., Green,
  A. J. \& Otrupcek, R. 1996, AJ, 111, 1651


\ref Gatley, I., Jones, T. J., Hyland, A. R., Beattie, D. H., \& Lee, T.
J. 1984, MNRAS, 210, 565

\ref Genzel, R., Hollenbach, D.J. \& Townes, C.H. 1994, 
Rep.Prog.Phys., 57, 417

\ref Ghez, A. et al. 1998, this conference

\ref Green, A.J., Frail, D.A., Goss, W.M. \& 
Otrupcek, R. 197, AJ, 114, 2058

\ref G\"usten et al. 1987, ApJ, 318, 124

\ref Jackson, J.M., Geis, N., Genzel, R., Harris, A.I.,   
  Madden, S., Poglitsch, A., Stacey, G.J., Townes, C.H. 1993, ApJ 402,  173

\ref Liszt, H. \& Burton, W.B. 1995, ApJS, 98, 679L

\ref Marr, J.M., Rudolph, A.L., Pauls, T.A., Wright, M.C., Backer, D.C. 1993,
 ApJ,  400, L29

\ref Melia, F., Fatuzzo, M., Yusef-Zadeh, F. \& Markoff, S. 1998a, ApJ,
508, L65

\ref Melia, F., Yusef-Zadeh, F. \& Fatuzzo, M. 1998b, ApJ, in press. 

\ref Menten, K.M., Goss, W.M., Ho, P.T.P. 1995, ApJ, 450, 1227

\ref Mezger, P.G., Duschl, W.J. \& Zylka, R. 1996, A.A.Rev., 7, 289

\ref  Morris, M. \& Serabyn, G. 1997, ARA\&A 34, 645

\ref Pak, S., Jaffe, D.T. \& Keller, L.D. 1996, ApJ, 457, L43

\ref Pauls, T., Johnston, K.J., Wilson, T.L., Marr, J.M. \& Rudolph, A. 1993,
ApJ, 403, L13

\ref  Pedlar, A., Anantharamaiah, K.R., Ekers, R.D., Goss, W.M.,
  van Gorkom, J.H., Schwarz, U.J., \& Zhao, J.-H. 1989, ApJ 342, 769

\ref Roberts, D. \& Goss, W.M. 1993, ApJS, 86, 133.

\ref  Serabyn, E., Lacy, J.H., \& Achtermann, J.M. 1992,
ApJ, 395, 166

\ref   Sternberg, A. \& Dalgarno, A. 1989, ApJ, 338, 197

\ref Telesco, C.M., Davidson, J.A. \& Werner, M.W. 1996, ApJ, 456, 541.

\ref  Wardle, M., Yusef-Zadeh, F., \& Geballe 1998, submitted to
ApJ.

\ref Wright, MCH, et al. 1987, in {\it The Galactic Center}, ed., DC
Backer, 133, New York: AIP

\ref Yusef-Zadeh, F., Lasenby, A.  \& Marshall, J. 1993, ApJ, 410, L27

\ref Yusef-Zadeh, F. \& Morris, M. 1987, ApJ, 320, 545

\ref Yusef-Zadeh, F., Roberts, D.A., Goss, W.M., Frail, D. \&
  Green, A. 1996, ApJ 466, L25

\ref Yusef-Zadeh, F., Roberts, D.A. \& Biretta, J. 
1998, ApJ 499, L159

  
\ref Yusef-Zadeh, F. \& Wardle, M. 1993, ApJ, 405, 584

\ref Zhao, J.H. \& Goss, W.M. 1998, ApJ, 499, L163

\ref Zhao, J.H., Goss, W.M., Ho, T.P.T. 1995 ApJ, 450, 1227

\ref Zylka, R., Mezger, P.G. \& Lesch, J. 1992, A.A. 261, 1197


\end{references}
\end{document}